\documentstyle[12pt]{article}
\centerline{\bf}
\newcommand{\be}{\begin{equation}}
\newcommand{\ee}{\end{equation}}
\newcommand{\bea}{\begin{eqnarray}}
\newcommand{\eea}{\end{eqnarray}}
\newcommand{\brr}{\begin{array}}
\newcommand{\err}{\end{array}}
\newcommand{\bc}{\begin{center}}
\newcommand{\ec}{\end{center}}
\newcommand{\nn}{\nonumber}
\newcommand{\GF}{\frac{G_{ F}}{\sqrt 2}}

\newcommand{\gammazeros}{\hat{\gamma}^{(0)T}_{ s}}
\newcommand{\gammazeroe}{\hat{\gamma}^{(0)T}_{ e}}
\newcommand{\gammaones}{\hat{\gamma}^{(1)T}_{ s}}
\newcommand{\gammaonee}{\hat{\gamma}^{(1)T}_{ e}}

\newcommand{\DSone}{\Delta S\!=\!1}
\newcommand{\DStwo}{\Delta S\!=\!2}

\newcommand{\Kzero}{K^{ 0}}
\newcommand{\Kbzero}{{\bar K}^{ 0}}

\newcommand{\md}{m_{ d}}
\newcommand{\ms}{m_{ s}}
\newcommand{\mc}{m_{ c}}
\newcommand{\mb}{m_{ b}}
\newcommand{\mt}{m_{ t}}
\newcommand{\Mw}{M_{ W}}

\newcommand{\Mwsq}{M_{ W}^{ 2}}

\newcommand{\J}{\hat{J}}
\newcommand{\Ke}{\hat{ K}}
\newcommand{\U}{\hat{U}}

\newcommand{\W}{\hat{     W}}

\newcommand{\PP}{\hat{ P}}

\newcommand{\Gmulup}{\gamma^{\mu}_{ L}}

\newcommand{\alphas}{\alpha_{ s}}
\newcommand{\alphae}{\alpha_{ e}}
\newfont{\bb}{msym10   scaled\magstep1}
\begin{document}
\title{$\epsilon^{\prime}/\epsilon$ at the next-to-leading order in
QCD and QED}
\author{ M. Ciuchini$^{1,2}$,
E. Franco$^2$, G. Martinelli$^2$ and L. Reina$^3$ \\
\\ $^1$ INFN, Sezione Sanit\`a, V.le Regina Elena 299,\\
00161 Roma, Italy,\\
$^2$ Dip. di Fisica ``G. Marconi",\\
Universit\`a degli Studi di Roma ``La Sapienza" and\\
INFN, Sezione di Roma, \\
P.le A. Moro 2, 00185 Roma, Italy,\\
$^3$ SISSA-ISAS, Via Beirut 2, 34014 Trieste and \\
Sezione INFN di Trieste, Via Valerio 2, 34100 Trieste, Italy \\}
\date{}
\maketitle
\begin{abstract}
We present a new calculation of the CP violation parameter $\epsilon^{\prime}/
\epsilon$. The results reported in this paper have been
obtained  by using  the $\Delta S=1$ effective Hamiltonian computed
 at the next-to-leading
order, including QCD and QED penguins. The matrix elements of the
relevant operators have been taken from lattice QCD, at a scale
$\mu=2$ GeV. At this relatively large scale,
the perturbative matching between the relevant operators and the
corresponding coefficients is quite reliable.\par  The effect of the
 next-to-leading
corrections is to lower the prediction obtained at the leading order,
thus favouring the experimental result of E731. We analyze  different
contributions to the final result and compare the leading
and next-to-leading cases.
 \end{abstract}
\vskip 5mm
\newpage
In this paper we present a theoretical prediction of $\epsilon^{\prime}/
\epsilon$ obtained from the effective weak hamiltonian ${\cal H}_{ eff}^
{\DSone}$, up to next-to-leading QCD and QED corrections. The Wilson
coefficients of the operators of ${\cal H}_{ eff}^{\DSone}$ have been
computed using the $(10\times 10)$ anomalous dimension matrix which governs
the mixing of the relevant current-current and penguin operators,
renormalized in the $\overline{MS}$ t'Hooft-Veltman
dimensional regularization scheme
(HV). The anomalous dimension matrix includes orders $(\alphas
t)^n$, $\alphas (\alphas  t)^n$, $\alphae t (\alphas t)^n$
and $\alphae (\alphas t)^n$ (where
$t=\ln\Mwsq/\mu^2$). The coefficients have been obtained by
integrating numerically the renormalization group equations from a scale
$\mu\!\sim\!\Mw\!\sim\!\mt$ down to a scale $\mu\!=\! 2\,\,\mbox{GeV}>\mc$.
The coefficients at the initial scale $\mu\!\sim\!\Mw$ are those computed
by Inami and Lim in ref. \cite{inami} and
by Flynn and Randall \cite{flynn1}, as corrected in ref. \cite{buras1}.
The initial effective
hamiltonian includes QCD, $Z^0$ and electro-magnetic penguins
and box diagrams. We have
neglected the running of the coefficients between $\mt$ and $\Mw$, but this
corresponds to a negligible error in the final result.

The matrix elements of the operators, expressed in terms of the
so-called  $B$-factors, have been taken, when possible, from lattice
calculations. It turns out that the most important $B$-factors have indeed
been determined on the lattice, with either Wilson and
staggered fermions \cite{shar3}-\cite{shar7}.
One notable exception is the $B$-factor of the operator $O^-$, believed to
be responsable for the $\Delta I= 1/2 $ enhancement.
For this reason we take the
non-leptonic $\Delta I= 1/2$ amplitude from experiments. For those matrix
elements not yet determined by lattice calculations, we allowed a variation
of the $B$-factor in the range $1$-$6$. The next-to-leading relation
between the operators in the lattice renormalization scheme and
$\overline{MS}$ naive dimensional regularization (NDR) is known
\cite{mart1}-\cite{mart2}
(the corrections are of the order of $30 \%$ in current lattice calculations)
and is usually included in the lattice results.
We have only computed at one loop the factors which relate the
operators renormalized in NDR to the corresponding ones in HV.
These corrections give a negligible shift to the values of the B-parameters.
Given the uncertainties in the evaluation of the B-parameters, we have
decided to use the same values as in ref. \cite{reina}, see Table \ref{Bpar}.
We have chosen the renormalization scale $\mu\!=\!
2\,\,\mbox{GeV}$, corresponding to the typical inverse lattice spacing at
which actual calculations of weak matrix elements are performed in lattice
QCD ($a^{-1}\!=2 \! -3.5\,\,\mbox{GeV}$).

There are great advantages in using weak matrix elements of operators from
lattice QCD. From the theoretical point of view, the matching of the
coefficients with the matrix elements of the operators, renormalized at the
scale $\mu$, is exact at the next-to-leading order in $\alphas$.
This is not the case with other methods.  For example, in the $1/N$
expansion, one has to match the coefficients, computed by renormalizing
the operators on quark states, with the matrix elements calculated in
the meson theory. However, it is not clear to us how this matching can be
 implemented at the next-to-leading order. \par
 A second
important point is that the scale $\mu$ can be taken as large as $2-
3\,\,\mbox{GeV}$, where the perturbative evaluation of the coefficients is
expected to be accurate and the final result quite stable for variations of
$\mu$ in the above range, see
below. This is to be contrasted with other approaches where
values of $\mu\sim 0.6-0.8\,\,\mbox{GeV}$ are chosen. At such low scales
the results are not stable against a variation of $\mu$. Moreover,
at low values of  $\mu$,
next-to-leading order (NLO)
corrections to the coefficients of penguin operators
are large, so that one can question the convergence of the perturbative
expansion.
\par The calculation of $\epsilon^{\prime}/\epsilon$ has been combined with
a next-to-leading order calculation of $\epsilon$ and of the $B^0$-$\bar B^0$
mixing amplitude, following the approach of ref. \cite{reina}.
 For these quantities
the next-to-leading perturbative corrections to the Wilson coefficients
have been known for quite a while and we do not have much to add to previous
 analyses.  We notice that, with the inclusion of the coefficients
and of the anomalous dimensions computed
at second order, the analysis of ref. \cite{reina}
is here consistently done at the next-to-leading accuracy.
The strategy is the same as  in ref. \cite{reina}:  from the comparison of the
theoretical value
of $\epsilon$ with the experimental number, given the uncertainties on
the matrix element of the $\Delta S=2$ operator and on the CKM parameters,
we find a range of allowed values of the CP violating phase $\delta$.
 Correspondingly
we compute $\epsilon^{\prime} / \epsilon$, which will also be affected
by a theoretical error, see Figs. (\ref{fig1})-(\ref{fig4}) below.
\par
The paper is organized as follows. We first give the
effective Hamiltonians  responsible for $\DStwo$ and
$\DSone$ transitions,
expressed in terms of the relevant operators and their Wilson coefficients.
{}From these  Hamiltonians we derive the expressions for
$\epsilon$ and
$\epsilon^{\prime}/\epsilon$, written as combinations
of the  Wilson coefficients
times the B-parameters of the different operators, i.e. their matrix elements.
We do not discuss the determination of the coefficients of the
operators relevant for $\epsilon$,
 since this point was explained in great detail in
ref. \cite{reina}.
We instead focus on the evaluation of the coefficients of the $\DSone$
Hamiltonian beyond the leading order including the effects of the
electro-weak penguins.  A comparison of the predictions in
the leading and next-to-leading cases is presented.
We also study the uncertainties in the final
evaluation of $\epsilon^{\prime}/\epsilon$ coming from the choice of the
scale and  $\Lambda_{QCD}$ and the dependence on the top
quark mass. Finally the theoretical predictions will be
confronted with the experimental results coming from NA31 \cite{na31} and
E731 \cite{e731}.  \par We have also considered the $B^0$-$\bar B^0$ transition
parameter $x_d$, following  the  analysis done in ref. \cite{reina},
but with a different range of $\Lambda^{n_f=4}_{QCD}=(340 \pm 120)$
MeV \cite{alto,german}.
It remains true
that a large value of $f_B$, for $m_t \ge 140$-$150$ GeV, favours a positive
value for  $\cos \delta $. Since the comparison of $x_d$ with the theoretical
 prediction has no other effect on
$\epsilon^{\prime} / \epsilon$ we will not
discuss it  any more.

\par
1) $\epsilon$ :
The effective Hamiltonian governing the $\DStwo$ amplitude is given by:
\bea
{\cal H}_{eff}^{|\Delta S|=2}
= \frac{G_{ F}^2}{16{\pi}^2}M_{ W}^2
({\bar d}{\gamma}^{\mu}_{ L}s)^2\left\{{\lambda}_c^2F(x_c)+{\lambda}_t
^2F(x_t)+2{\lambda}_c{\lambda}_tF(x_c,x_t)\right\} \nonumber \\
& &
\label{effective hamiltonian1}
\eea
where $G_{ F}$ is the Fermi coupling constant and
$\gamma^{\mu}_{ L}=\gamma^{\mu}(1-\gamma_5)\,$;
${\lambda}_q$'s are related to the CKM matrix elements by
${\lambda}_q=V^{\star}_{qi}V_{qf}$,
where `$i$' and `$f$' are the labels of the initial and final states
respectively (in the present case
$i=s$ and $f=d$).
$x_q={m_q^2}/{M_{ W}^2}~$ and
the functions $F(x_i)$ and $F(x_i,x_j)$ are the so-called
{\it Inami-Lim} functions \cite{inami}, obtained from the calculation of the
basic box-diagram and including QCD corrections \cite{buras0}.
 $F(x_t)$ is known at the
next-to-leading order, which has been included in our calculation.
{}From eq. (\ref{effective hamiltonian1}) we can derive the CP violation
parameter $\epsilon$:
\be
|\epsilon|_{\xi=0}=C_{ \epsilon}B_{ K}A^2\lambda^6\hat \sigma
\left\{F(x_c,x_t)+F(x_t)[A^2\lambda^4(1-\rho)]
-F(x_c)\right\}
\label{epsilon_csizero_1}
\ee
where
\be
C_{ \epsilon}=\frac
{G_{ F}^2f_{ K}^2 m_{ K}M_{ W}^2}{6\sqrt 2{\pi}^2\Delta M_K}
\ee
$\Delta M_K$ is the mass difference between the two neutral kaon mass
eigenstates.
In eq. (\ref{epsilon_csizero_1}) $\hat \sigma = \sqrt{ \rho^2 +\eta^2}
\sin \delta$ and $A$, $\rho$, $\eta$ and $\delta$ are the parameters of
the CKM matrix in the Wolfenstein  parametrization \cite{wolf}.
$B_{ K}$ is the renormalization group invariant
B-factor, defined as :
\be
B_{ K}=B_{ K}(\mu)\left[\alpha_{ QCD}(\mu)\right]^{-6/25}
\label{bkfactor}
\ee
$B_K$ takes into account all
the possible deviations from the vacuum insertion approximation in the
evaluation of the $\langle\Kbzero|(\bar d\Gmulup s)^{ 2}|\Kzero\rangle$
matrix element ($B_{ K}=1$ corresponding to an exact vacuum insertion
approximation).
\par 2) $\epsilon^{\prime}/\epsilon$:
Most of the discussion in this paper is devoted to
the Wilson coefficients of the operators
appearing in the effective $\Delta S=1$ Hamiltonian, which we have
computed at the next-to-leading order, including QCD and
QED corrections. \par
The $\Delta S=1$ effective hamiltonian is given by:
\be
{\cal H}_{eff}^{|\Delta S=1|}=\sum_{i}C_{ i}(\mu)Q_{ i}(\mu)
\label{eh} \ee
The complete basis of operators when QCD and QED corrections are taken
into account is given by:
\bea
Q_{ 1}&=&({\bar s}_{\alpha}d_{\alpha})_{ (V-A)}
    ({\bar u}_{\beta}u_{\beta})_{ (V-A)}
   \nn\\
Q_{ 2}&=&({\bar s}_{\alpha}d_{\beta})_{ (V-A)}
    ({\bar u}_{\beta}u_{\alpha})_{ (V-A)}
\nn \\
Q_{ 3,5} &=& ({\bar s}_{\alpha}d_{\alpha})_{ (V-A)}
    \sum_{q=u,d,s}({\bar q}_{\beta}q_{\beta})_{ (V\mp A)}
\nn \\
Q_{ 4,6} &=& ({\bar s}_{\alpha}d_{\beta})_{ (V-A)}
    \sum_{q=u,d,s}({\bar q}_{\beta}q_{\alpha})_{ (V\mp A)}
\nn \\
Q_{ 7,9} &=& \frac{3}{2}({\bar s}_{\alpha}d_{\alpha})_
    { (V-A)}\sum_{q=u,d,s}e_{ q}({\bar q}_{\beta}q_{\beta})_
    { (V\pm A)}
\nn \\
Q_{ 8,10} &=& \frac{3}{2}({\bar s}_{\alpha}d_{\beta})_
    { (V-A)}\sum_{q=u,d,s}e_{ q}({\bar q}_{\beta}q_{\alpha})_
    { (V\pm A)} \nn \\
Q^c_{ 1}&=&({\bar s}_{\alpha}d_{\alpha})_{ (V-A)}
    ({\bar c}_{\beta}c_{\beta})_{ (V-A)}
\nn \\
Q^c_{ 2}&=&({\bar s}_{\alpha}d_{\beta})_{ (V-A)}
    ({\bar c}_{\beta}c_{\alpha})_{ (V-A)}
\label{epsilonprime_basis}
\eea
where the subscript $(V \pm A)$ indicates the chiral structure and
$\alpha$ and $\beta$ are colour indices.
\par The operators $Q_i(\mu)$ are renormalized at the scale $\mu$ in
$\overline{MS}$ , using the HV
regularization scheme.  The corresponding coefficients,
$C_i(\mu)$ are scheme dependent. The dependence on the regularization
scheme appears at one loop, when we express the original current-current
product
in terms of the Wilson OPE:
\bea < F \vert {\cal H}_{eff}^{|\Delta S=1|}\vert I > &=& g_W^2/8 \int d^4 x
D_W(x^2, M_W^2) < F \vert T \left( J_{\mu}(x),J^{\dagger}_{\mu} (0) \right)
\vert I > \nn \\ &\rightarrow& \sum_{i}
C_{ i}(M_W) < F \vert Q_{ i}(M_W)  \vert I > \eea
The Wilson coefficients $\vec C(M_W)=(C_1(M_W), C_2(M_W) ...)$
are found by matching, at $O(\alphae)$
and $O(\alphas)$ in HV, the current-current and
penguin diagrams computed with the $W$ and top propagators to
those computed with the local four-fermion operators in the effective theory.
\par
$\vec C(\mu)$ are expressed
in terms of  $\vec C(M_W)$ through the
 renormalization evolution matrix $\W[\mu,M_W]$\footnote{ We have
properly taken into account the beauty threshold in the evolution matrix.}:
\be \vec C(\mu) = \W[\mu,M_W] \vec C(M_W) \label{evo} \ee
where:
\bea
\W[\mu,M_W]   = \hat M[\mu] \U[\mu, M_W] \hat M^{\prime}[M_W]
 \label{monster} \eea
with:
\be  \hat M[\mu] =
\left(\hat 1 +\frac{\alphae }{4\pi}\Ke\right)
  \left(\hat 1 +\frac{\alphas (\mu)}{4\pi}\J\right)
 \left(\hat 1+\frac{\alphae}{\alphas (\mu)}\PP\right)
\label{mo1} \ee
and
\be \hat M^{\prime}[M_W]=\left(\hat 1-\frac{\alphae}{\alphas (M_W)}\PP\right)
            \left(\hat 1 -\frac{\alphas (M_W)}{4\pi}\J\right)
\left(\hat 1 -\frac{\alphae }{4\pi}\Ke\right)
\label{mo2} \ee

At the next-to-leading accuracy $\W[\mu,M_W]$ is regularization scheme
dependent.
\par Eqs. (\ref{monster}), (\ref{mo1}) and
(\ref{mo2}) require a detailed explanation.
At the leading order, the QCD anomalous dimension matrix,
including QCD penguins, has been
computed in refs. \cite{russi,gilman}. The electro-weak anomalous dimension
matrix at the same order can be found in refs. \cite{buras1}, \cite{bij}
and \cite{lus}.
We have computed
the anomalous dimension matrix at the next-to-leading order,
  by calculating all the current-current and penguin
operators at two loops up to order $\alphas^2 t$  and $\alphae \alphas t$.
This corresponds to all the diagrams with four external
quark legs, where one of the operators in the list given
in eqs. (\ref{epsilonprime_basis})
is inserted and two gluons or one gluon and one  photon are exchanged.
At $O(\alphas^2 t)$,
the two loop anomalous dimension matrix was computed in refs. \cite{alta,bur3}
for current-current diagrams and in ref. \cite{bur4} for penguin diagrams.
The explicit expression of the anomalous dimension matrix alone
would take more space than that allowed for a letter and will
be presented in a separate publication \cite{lungo}.
Here we simply explain the
meaning of all  the terms appearing in eq. (\ref{monster})-(\ref{mo2}).
To obtain the expression in eq. (\ref{monster})-(\ref{mo2})
 we have neglected the
running of the coefficients between the top quark mass and the
$W$ mass. We have also expanded the formula at first order in $\alphae$ and
neglected the running of the electro-magnetic coupling. These approximations
are immaterial for the final numerical result.
\par The matrix $\U[\mu,M_W]$ in eq. (\ref{monster}) is given by
\be
\U[\mu,M_W]= \left[\frac{\alphas (M_W)}{\alphas (\mu)}\right]^{
            \gammazeros / 2\beta_{ 0}}
\label{u0} \ee
The matrices $\PP$, $\J$ and $\Ke$ are solutions of the equations:
\be
\PP+\left[\PP,\frac{\gammazeros}{2\beta_{ 0}}\right] = \frac{\gammazeroe}
   {2\beta_{ 0}} \label{pp} \ee
\be
\J-\left[\J,\frac{\gammazeros}{2\beta_{ 0}}\right] =
         \frac{\beta_{ 1}}{2\beta^2_{ 0}}\gammazeros-
         \frac{\gammaones}{2\beta_{ 0}} \label{jj}
\ee
\be
\left[\Ke,\gammazeros \right]
=\gammaonee+\gammazeroe \J+\gammaones \PP+\left[\gammazeros,\J\PP\right]
    -2\beta_1 \PP -\frac{\beta_{ 1}}{2\beta_{ 0}}\PP\gammazeros
\label{ke} \ee
The anomalous dimension matrix, which includes gluon and photon corrections
has been separated in several pieces which appear in the above equations:
\be \hat \gamma= \frac {\alphas }{ 4 \pi } \hat \gamma_s^{(0)} +
 \frac {\alphae }{4 \pi} \hat \gamma_e^{(0)}
+ (\frac {\alphas }{4 \pi})^2 \hat \gamma_s^{(1)} +
 \frac{ \alphas }{4 \pi} \frac{ \alphae}{4 \pi}  \hat \gamma_e^{(1)}
\nn \ee
where each of the $\hat \gamma^{(0,1)}_{s,e}$ is a $10 \times 10$ matrix.
In
eqs. (\ref{u0}-\ref{ke}), $\beta_{ 0}$ and $\beta_{ 1}$ are the
first two coefficients of the $\beta$-function of $\alphas$.
\par From ${\cal H}_{eff}^{|\Delta S=1|}$
we can derive the expression for $\epsilon^{\prime}$:
\be  \epsilon^{\prime}=\frac{e^{ i\pi/4}}{\sqrt{2}}\frac{\omega}
{\mbox{Re}A_{ 0}}\left[\omega^{ -1}
(\mbox{Im}A_{ 2})^{\prime}-(1-\Omega_{ IB})\,\mbox{Im}A_{ 0}
\right]
\label{epsilonprime}
\ee
where
$(\mbox{Im}A_{ 2})^{\prime}$ and $\mbox{Im}A_{ 0}$ are given by:
\bea
\mbox{Im}A_{ 0} &=&-\GF Im\Bigl({ V}_{ ts}^{ *}{ V}_{ td}\Bigr)
\left\{-\left(C_{ 6}B_{ 6}+\frac{1}{3}C_{ 5}B_{ 5}\right)Z
+\left(C_{ 4}B_{ 4}+\frac{1}{3}C_{ 3}B_{ 3}\right)X+\right.
\nn\\
& &\!C_{ 7}B_{ 7}^{ 1/2}\left(\frac{2Y}{3}+\frac{Z}{6}-
\frac{X}{2}\right)+C_{ 8}B_{ 8}^{ 1/2}\left(2Y+\frac{Z}{2}+
\frac{X}{6}\right)-\nn\\
& &\!\left.C_{ 9}B_{ 9}^{ 1/2}\frac{X}{3}+\left(\frac{C_{ 1}
B_{ 1}^{ c}}{3}+C_{ 2}B_{ 2}^{ c}\right)X\right\}
\label{ima0}
\eea
and
\bea
(\mbox{Im}A_{ 2})^{\prime}\!&=&\!-G_{ F}Im\Bigl({ V}_{ ts}^{ *}{ V}_{ td}\Bigr)
\left\{C_{ 7}B_{ 7}^{ 3/2}\left(\frac{Y}{3}-\frac{X}{2}\right)+
\right. \nn \\
& &\!\left.C_{ 8}B_{ 8}^{ 3/2}\left(Y-\frac{X}{6}\right)+
C_{ 9}B_{ 9}^{ 3/2}\frac{2X}{3}\right\}
\label{ima2}
\eea
$ \omega=\mbox{Re}A_{ 2}/ \mbox{Re}A_{ 0}=0.045 $ and
we have introduced $( \mbox{Im}A_{ 2})^{\prime}$ defined as:
\be
\mbox{Im}A_{ 2}=(\mbox{Im}A_{ 2})^{\prime}+\Omega_{ IB}
(\omega\,\mbox{Im}A_{ 0})
\ee
$\Omega_{ IB}= +0.25 \pm 0.10$
represents the isospin breaking contribution, see for example ref. \cite{bur5}.
\par The numerical evaluation of $\epsilon^{\prime}/\epsilon$ requires the
knowledge  of the Wilson coefficients of the operators and of the
corresponding matrix elements.
The Wilson coefficients have been
evaluated,  using eq. (\ref{evo}), combined with the evolution matrix
of eq. (\ref{monster}), and the initial conditions computed in
refs. \cite{flynn1,buras1} (and given for HV in ref. \cite{bur4}).
The matrix elements of the operators have been written in terms of the
three quantities (see eqs. (\ref{ima0}) and
(\ref{ima2})) :
\bea
X\!&=&\!f_{\pi}\left(M_{ K}^{ 2}-M_{\pi}^{ 2}\right)\\
Y\!&=&\!f_{\pi}\left(\frac{M_{ K}^{ 2}}{\ms(\mu)+\md(\mu)}\right)
^2\sim 12\,X\left(\frac{0.15 \, \mbox{GeV}}{\ms(\mu)}\right)^2\\
Z\!&=&\!4\left(\frac{f_{ K}}{f_{\pi}}-1\right)Y
\eea
and a set $\{B_{ i}\}$ of B-parameters (in our normalization $f_{\pi}
=132$ MeV).
The numerical value of the B-parameters have been taken from lattice
calculations and multiplied by suitable renormalization factors
to take into account the difference between HV and the lattice regularization
scheme. For those B-factors which have not been computed yet on the lattice
we have used an educated guess, which will be discussed below.
We observe that in eqs. (\ref{ima0}) and (\ref{ima2}) only nine
coefficients (B-parameters) appear since we have used the relation
$ Q_{ 10}=-Q_{ 3} + Q_{ 4} + Q_{ 9}$.
\begin{table}
\begin{center}
\begin{tabular}{c c c c c c}\hline\hline\\
$B_{ K},B_{ 9}^{ (3/2)}$ &  $B_{ 1-2}^{ c}$ &
 $B_{ 3,4}$ &
$B_{ 5,6}$ & $B_{ 7-8-9}^{ (1/2)}$ & $B_{ 7-8}^{ (3/2)}$
\\ \\\hline \\
$0.8\pm 0.2$ &  $0 - 0.15^{ (*)}$ & $1 -  6^{ (*)}$ &
$1.0\pm 0.2$ & $1^{ (*)}$ & $1.0\pm0.2$
\\ \hline\hline
\end{tabular}
\caption[]{Values of the $B$-parameters. Entries with a $^{ (*)}$
are educated guesses; the others are taken from lattice QCD calculations.}
\label{Bpar}
\end{center}
\end{table}
\par We will call ``central"  results obtained by using
the central values of the B-parameters reported in Table \ref{Bpar}, $\mu = 2$
GeV and $\Lambda_{QCD}=340$ MeV (4 flavours).
We will  study the uncertainty of the
theoretical prediction by varying the B-parameters, $m_s$ and
the experimental quantities, such as  $\Lambda_{QCD}$,
the CKM mixing parameters $A$, $\rho$, etc.,
in the range indicated by the
errors on the quantities reported in Tables 1 and 2.
In the study of the differences between $\epsilon^{\prime}/\epsilon$,
computed at the leading order (LO) or next-to-leading order (NLO), we
will keep fixed the coefficients at the scale $M_W$. This
means that, also at the LO, the coefficients include
the $O(\alphas)$ and $O(\alphae)$  corrections which arise from the matching
between the effective hamiltonian and the original current-current product.
We also keep fixed $\Lambda_{QCD}$  in passing from the LO
calculation to the NLO one.
\begin{figure}[t]   
    \begin{center}
       \setlength{\unitlength}{1truecm}
       \begin{picture}(6.0,6.0)
          \put(-6.0,-6.2){\special{nfig2_nlo100_r.ps}}
       \end{picture}
    \end{center}
    \caption[]{ Band of allowed
values for $\epsilon^{\prime}/\epsilon$ at  $m_t=100$ GeV (next-to-leading
order). The dashed lines represent the experimental
results of NA31, $(2.3 \pm 0.7)\cdot 10^{-3}$ and E731, $(0.6 \pm 0.7 )
\cdot 10^{-3}$.}
    \protect\label{fig1}
\end{figure}
\begin{figure}[t]   
    \begin{center}
       \setlength{\unitlength}{1truecm}
       \begin{picture}(6.0,6.0)
          \put(-6.0,-6.2){\special{nfig2_nlo140_r.ps}}
       \end{picture}
    \end{center}
    \caption[]{Same as in Fig.(\ref{fig1}) at  $m_t=140$ GeV.}
    \protect\label{fig2}
\end{figure}
\begin{figure}[t]   
    \begin{center}
       \setlength{\unitlength}{1truecm}
       \begin{picture}(6.0,6.0)
          \put(-6.0,-6.2){\special{nfig2_nlo200_r.ps}}
       \end{picture}
    \end{center}
    \caption[]{Same as in Fig.(\ref{fig1}) at  $m_t=200$ GeV.}
    \protect\label{fig3}
\end{figure}
\begin{figure}[t]   
    \begin{center}
       \setlength{\unitlength}{1truecm}
       \begin{picture}(6.0,6.0)
          \put(-6.0,-6.2){\special{nfig2_llo100_r.ps}}
       \end{picture}
    \end{center}
    \caption[]{ Same as in Fig.(\ref{fig1}) at  $m_t=100$ GeV, with the
coefficients of the operators for $\epsilon^{\prime}/\epsilon$
computed at the leading order.}
    \protect\label{fig4}
\end{figure}
\begin{table}
\begin{center}
\begin{tabular}{c c}\hline
{\it parameter} & {\it
value} \\ \hline \\
$\Lambda_{ QCD}$ & $340\pm 120$ GeV\\
$\ms(2\,\mbox{GeV})$ & $(170\pm 30)$ MeV\\
$\mc(2\,\mbox{GeV})$ & $1.5$ GeV \\
$\mb(2\,\mbox{GeV})$ & $4.5$ GeV \\
$A\lambda^2$ & $0.047\pm 0.004$\\
$\sqrt{\rho^2+\eta^2}={ V}_{ ub}/(\lambda{ V}_{ cb})$ & $0.50\pm 0.14$\\
$\epsilon_{exp}$ & $2.28\cdot 10^{-3}$\\
$\mbox{Re}A_{ 0}$ & $2.7\cdot10^{-7}\mbox{GeV}$\\
\hline
\end{tabular}
\caption[]{Values of experimental parameters used in this work.}
\label{tab_val}
\end{center}
\end{table}

\par We first consider the
relative contribution of different operators to
$\epsilon^{\prime}/\epsilon$. Following the standard notation we write:
\be  \epsilon^{\prime}/\epsilon  \sim  R   \times
 C_{ 6} B_{ 6} \Bigl( 1- \sum_i \Omega_{ i} \Bigr) \label{kfact} \ee
where  the coefficient $R$ includes $\sin \delta$. In eq.(\ref{kfact})
$\Omega_{ i}=\Omega_{ 4}, \Omega_{ 8}^{3/2}$, etc. indicate
the relative corrections due to the matrix elements of the operators,
different from $Q_6$, appearing in eqs. (\ref{ima0})-(\ref{ima2}).
For example, the central value of $R$, averaged over the
values of $\cos \delta$ allowed by the analysis of $\epsilon$, is $\vert
R \vert \sim
1.0\cdot 10^{-2}$  for $m_t=140$ GeV. \par
In the following discussion we fix $\mu=2$ GeV and $m_t= 140$ GeV
and we allow a variation of the B-parameters, computed on the
lattice,
around  their central values. For those matrix elements
still to be computed we proceed as follows. We fix
$B^{1/2}_{7-9}$ equal to one,
since their contribution  is of the order of few percent.
If we take $B_{ 3, 4}=1$  and $B^c_{1,2}=0$,
at the LO the largest contributions to $\epsilon^{\prime}/\epsilon$
come from $Q_{ 6}$, $Q_{ 8}$ and $Q_{ 9}$,
$\Omega_{ 8}^{3/2} \sim 28 \%$ and
$\Omega_{ 9}^{3/2}  \sim -18 \%$.
 $C_{ 9}$ is much larger than $C_{ 8} $ and
compensates for the matrix element of
$Q_{ 9}$ which is much smaller than
the matrix element of $Q_{ 8} $. With $B_{ 3,  4}=1$, $\Omega_{ 3}
\sim -0.01$ and $\Omega_{ 4} \sim 0.04$.
 The operators $Q_{ 3 , 4}$ have a chiral structure similar
to $Q_{ 2}$ and, at scales larger than $m_c$, give rise to the same
``eye" diagrams, like $Q_{ 2}$ does. In order to explain the
experimental $\Delta I=1/2$ enhancement, $B_{ 2}$ must be of the
order $5$-$6$. For this reason in ref. \cite{reina} and in the present
calculation we have allowed a variation of
$B_{ 3,  4}$ between 1 and 6. The central values in this case
are $\Omega_{ 3}
\sim -0.04$ and $\Omega_{ 4} \sim 0.28$. Thus, if $Q_{ 4}$
has a large B-factor, its contribution is as important as the
contribution of  $Q_{ 8}$ or  $Q_{ 9}$.
 $B_{ 1 , 2}^c$ are zero in the vacuum saturation approximation.
Their coefficients however are so large  that even a small B-factor can give
a sizeable contribution to $\epsilon^{\prime}/\epsilon$. By varying
$B_{ 1 , 2}^c$ between $0$-$0.15$ we find that $
\Omega_{ 2}^c \sim -0.18$ ($\Omega_{ 1}^c \sim 0.02$),
comparable to other large terms.
\begin{figure}[t]   
    \begin{center}
       \setlength{\unitlength}{1truecm}
       \begin{picture}(6.0,6.0)
          \put(-4.5,-6.2){\special{ncf68.ps}}
       \end{picture}
    \end{center}
    \caption[]{ $C_6$ and $C_8$ as a function of $\mu$  for $\Lambda_{QCD}=220$
(dotted),$340$ (solid) and $460$ (dashed) MeV. }
    \protect\label{fig5}
\end{figure}
\begin{figure}[t]   
    \begin{center}
       \setlength{\unitlength}{1truecm}
       \begin{picture}(6.0,6.0)
          \put(-6.0,-6.2){\special{ncf8lnl.ps}}
       \end{picture}
    \end{center}
    \caption[]{ $C_8$ as a function of $\mu$  for $\Lambda_{QCD}=340$
MeV at the LO and NLO. }
    \protect\label{fig6}
\end{figure}

  In the following we will focus on the main contributions
to $\epsilon^{\prime}/\epsilon$, which are due to few operators,
$Q_{ 4,  6, 7, 8,  9}$ and
$Q^c_{ 2}$. All the other terms are of the order
of a few $\%$,  with alternating signs and we will not discuss them any more.
 \par We now come to the effects
 of the next-to-leading corrections.
If we only include next-to-leading corrections due to two gluon
exchanges (corresponding to $\hat \gamma^{(1)}_s$), all the coefficients
are slightly changed, without modifying the pattern of the different
contributions observed at the leading order. For example
$\Omega_{ 4}$ goes from $0.28$ to $0.34$,
$\Omega_{ 5}$ from $0.06$ to $0.08$ and
$\Omega_{ 8}^{3/2}$ from $0.28$ to $0.26$.
 $\Omega_{ 9}^{3/2}$ increases  from
$-0.18$ to $-0.22$. The gluon-photon corrections on $\Omega_{ i}$
are always very tiny and,
given the uncertainties on the B-factors, practically invisible.
The only exceptions are $\Omega_{ 7, 8}^{3/2}$:
$\Omega_{ 7}^{3/2}$ is very
small at the leading order, $-0.01$, becomes  larger including two-gluon
exchanges, $-0.05$ and becomes $-0.09$  with
 the  complete next-to-leading corrections;
 $\Omega_{ 8}^{3/2}$ is $0.28$ at the LO and
 increases to $0.51$ with the
complete NLO corrections. We notice that the sum of the contributions
of $Q_{ 7,  8}$ and $Q_{ 9}$ almost cancel at the leading, as
well as at the next to leading order. $\Omega_{ 7}^{3/2}+
\Omega_{ 8}^{3/2}+\Omega_{ 9}^{3/2} \sim 0.10$ at the leading order and
$0.20$ at the next-to-leading one. Similarly $\Omega_{ 4}+
\Omega^c_{ 2} \sim +10 \%$ both at the LO and at NLO.
We thus find that at
$m_t=140$ GeV and for the ``central"
values of the B-parameters,  the result is essentially
the same as the original Gilman-Wise prediction, since the most important
electro-penguin corrections almost cancel and their sum is
of the order of a few per cent\footnote{
The cancellation between different contributions may not occur for values of
the
B-parameters different from their central ones. This will correspond to a band
of uncertainty in  the theoretical prediction for $\epsilon^{\prime}/
\epsilon$, which is reported in the Figs. (\ref{fig1})-(\ref{fig4}).}.
  Thus $\epsilon^{\prime}/
\epsilon$ is essentially determined by $C_{ 6}$ and the corresponding
B-parameter, $B_{ 6}$.
$C_{ 6}$  varies from $-6.2 \times 10^{-2}$ to $-5.1 \times 10^{-2}$, at
fixed $\Lambda_{QCD}=340$ MeV, so that the average value of $\epsilon^{\prime}/
\epsilon$ decreases from $6.5 \times 10^{-4}$ ( LO)  to $4.5 \times 10^{-4}$
(NLO). The decrease of $C_6$ is mainly due to the variation
of the value of $\alpha_s$ between LO and NLO, if we insist in using
the same value of $\Lambda_{QCD}$. The relative uncertainty
on $C_6$, by using different
values of $\Lambda_{QCD}$,  remains roughly the same between
the LO, $\delta C_6/C_6 \sim 18\%$ and NLO,  $\delta C_6/C_6 \sim 17\%$.
\par As observed already in ref. \cite{flynn1}, and confirmed by
all other analyses \cite{buras1,reina}, for increasing values of the top mass,
 the
contribution of the electro-penguin operators tends to cancel the
contribution coming from $Q_6$. We have already observed that
at $m_t=140$ GeV  the main corrections come from $\Omega^{3/2}_{7,8,9}$,
$\Omega_4$ and $\Omega_2^c$. $\Omega_4$ and $\Omega_2^c$ vary very
little with $m_t$ and their values remain essentially the same in
going from LO to NLO. We know very little
about their matrix elements from non-perturbative calculations. We estimate
that the uncertainty on the final result, coming from the poor knowledge
of the matrix elements of $Q_4$ and $Q_2^c$ is of the order
of $30 \%$. On the other hand $\Omega^{3/2}_{7,8,9}$ vary  with $m_t$
and change considerably from LO to NLO. For example, at
$m_t=140$ GeV  $\Omega^{3/2}_8$ is $\sim 0.28$ at LO and $\sim 0.50$ at NLO,
at $m_t=200$ GeV it becomes $0.68$ at LO and $0.90$ at NLO.
We also observe large variations for $\Omega^{3/2}_7$ and $\Omega^{3/2}_9$
\footnote{ The increase of the relative contribution of
$\Omega^{3/2}_{7-9}$ is  due to the decrease of $C_6$ combined with
an increase of $C_{7-9}$ between the leading and next-to-leading cases.}.
The sum $\Omega^{3/2}_7+\Omega^{3/2}_8+\Omega^{3/2}_9 $
is $\sim 20 \%$ at $m_t=140$ GeV and
$\sim 65 \%$ at $m_t=200$ GeV. The same can be said also at low values
of the top mass, for example $m_t=100$ GeV.
In summary, in the range of $m_t$ considered
in this work, even though single contributions from the electro-penguin
operators may change by $\sim 40 \%$
because of the next-to-leading corrections, they give globally
more or less the same relative contribution at NLO as at the leading order.
Since for fixed  $\Lambda_{QCD}$, $C_6$ decreases, the net effect is the
the central value of the theoretical prediction for $\epsilon^{\prime}
/ \epsilon$ is smaller at the NLO.
In Figs. (\ref{fig1})-(\ref{fig3}), we report our results at NLO for $\epsilon^
{\prime}/ \epsilon$, at three different values of $m_t$, $100$, $140$ and
$200$ GeV. The predictions are given for different values of $\cos \delta$,
which are compatible with the analysis of $\epsilon$ \cite{reina}.
They have been obtained by comparing
the theoretical expressions for $\epsilon$,
eqs. (\ref{epsilon_csizero_1}), to the corresponding experimental value.
The dashed lines indicate the results by the NA31 and E731 experiments
at the 1-$\sigma$ level. The numbers and errors reported in the figures
are the average and the theoretical variance (for $\cos \delta$ positive
or negative) computed
by varying the B-parameters and the other quantities in the ranges
reported in Table \ref{Bpar} and \ref{tab_val}.
Since the value of $\epsilon^{\prime}
/ \epsilon$ is lowered at the next-to-leading order, our theoretical
prediction, at $m_t=140$ GeV,
 is now centered on the experimental result of E731 ($\cos
\delta \ge 0$)\footnote{ We recall that the most recent
lattice results \cite{allton}-\cite{sacsac}
 and QCD sum rules calculations \cite{neu}-\cite{rbo} suggest a
large value of $f_B$ which corresponds to $\cos \delta
\ge 0$ when $m_t \ge 140$-$150$ GeV \cite{reina}.}. For a comparison
between the LO and the NLO, we report
in Fig. (\ref{fig4}) the LO result, at $m_t=100$
GeV. The LO result in this case
 sits in the middle between the measurements of NA31 and
E731\footnote{The corresponding figure
in ref. \cite{reina} was obtained with different values of
$\Lambda_{QCD}$.}.
We also observe that the band of error is slightly reduced by the
inclusion of the next-to-leading corrections. Indeed the relative
error on $\epsilon^{\prime}
/ \epsilon$  at LO and NLO is basically the same. Since the
central value of $\epsilon^{\prime}
/ \epsilon$ is decreased at NLO, the band in Fig. (\ref{fig1}) appears narrower
than in Fig. (\ref{fig4}).
\par Before concluding this paper we want to discuss the stability of
the Wilson coefficients, i.e.
how much their values are affected by the presence of
NLO corrections and their sensitivity with respect to a variation of $\mu$ and
$\Lambda_{QCD}$. We observe that for $\mu \le 1$ GeV the values of the
coefficients start to vary wildly if one changes $\Lambda_{QCD}$ or
$\mu$. As an example, in Fig. (\ref{fig5}), we report the variation
of $C_6(\mu)$ and $C_8(\mu)$ as a function of $\mu$, at three different values
of
$\Lambda_{QCD}$ and $m_t=140$ GeV. The coefficients change by a factor of 2 or
more,
for $\mu \le 1$ GeV, if we vary $\Lambda_{QCD}$ from $220$ to $460$
MeV. Moreover, at small values of $\mu$,
the difference between the leading and next-to-leading
results is very large, and we cannot trust the perturbative
expansion. To ilustrate this point, we report in Fig. (\ref{fig6})
$C_8(\mu)$, for $0.8$ GeV $\le
\mu \le$ $4.0$ GeV at the leading  and next-to-leading order.
On the basis of the above discussion
we believe that a realistic  prediction for
$\epsilon^{\prime}/ \epsilon$ can only be obtained by matching
the Wilson coefficients and the operators matrix elements at scales
larger than $m_c$, as one can do by taking the matrix elements from
 lattice QCD.
\par The conclusions of this work are the following. It is reassuring
that by taking a renormalization scale $\mu$ of the order of $2$ GeV,
the Wilson coefficients of the effective Hamiltonian do not vary by
more than $\sim 35 \%$ and that the final result on
$\epsilon^{\prime}/\epsilon$ is quite stable in going from the leading to
the next-to-leading order. At NLO one gives a precise meaning to
the value of $\Lambda_{QCD}$ to be used. It is now consistent
to take  $\Lambda_{QCD}$ from the measurements done  in deep inelastic
scattering or at LEP.  The NLO calculation of the anomalous dimension
matrix makes also
consistent the evolution of the Wilson
coefficients with their matching at
$\mu \sim m_t\,,\,M_W$. Indeed the matching procedure
 is a next-to-leading order
effect. Finally, in the operator product expansion, only the NLO calculation
fixes  unambigously  the scale at which the operators must be renormalized.
In the lattice case the scale is dictated, up to higher order  effects,
by  the inverse
lattice spacing at which the numerical simulations are performed. The major
source of theoretical uncertainty remains now the evaluation
of the matrix elements, which are determined with large errors or are still
to be determined, like it is the case for $O^-$. \par
{}From the phenomenological point of view, at fixed $\Lambda_{QCD}$ and
$m_t$, the
next-to-leading corrections  lower the theoretical predictions, thus
favouring the experimental result by E731. \vskip 0.5 cm
\centerline{Acknowledgments}
\par We thank P. Nason for his participation to the earlier stages
of this work. We also thank G. Altarelli,
A. Buras, M. Lusignoli, L. Maiani, G. Salina and P. Weisz for many
interesting discussions on the subject of this paper.

\end{document}